\documentstyle[psfig,aps,prb,eqsecnum,tighten,floats]{revtex}

\psdraft

\begin{document}
\draft

\twocolumn[\hsize\textwidth\columnwidth\hsize
\csname@twocolumnfalse\endcsname

\title{The two-dimensional $t$-$t^{\prime }$-$U$ model as a minimal model for cuprate materials}
\author{Adolfo Avella, Ferdinando Mancini and Dario Villani}
\address{Dipartimento di Scienze Fisiche {\em ''E.R. Caianiello''} - Unit\`a INFM di Salerno\\
Universit\`a degli Studi di Salerno, 84081 Baronissi (SA), Italy}
\author{Hideki Matsumoto}
\address{Department of Applied Physics, Seikei University, Tokyo 180, Japan}
\date{\today}
\maketitle

\begin{abstract} \widetext
The addition to the Hubbard Hamiltonian of a $t^{\prime }$ diagonal hopping term, which is considered to be material dependent for high-$T_c$ cuprate superconductors, is generally suggested to obtain a model capable to describe the physics of high-$T_c$ cuprate materials. In this line of thinking, the two-dimensional $t$-$t^{\prime }$-$U$ model has been studied by means of the Composite Operator Method, which allows to determine the dynamics in a fully self-consistent way by use of symmetry requirements, as the ones coming from the Pauli principle. At first, some local quantities have been calculated to be compared with quantum Monte Carlo data. Then, the structure of the energy bands, the shape of the Fermi surface and the position of the van Hove singularity have been computed as functions of the model parameters and studied by the light of the available experimental data. The results of our study show that there exists two sets of parameters that allows the model to describe the relevant features of $1$-layer compounds $Nd_{2-x}Ce_xCuO_4$ and $La_{2-x}Sr_xCuO_4$. On the other hand, for the $2$-layer compound $YBa_2Cu_3O_{7-\delta }$ is not possible to find a reasonable set of parameters which could reproduce the position of the van Hove singularity as predicted by {\em ARPES} experiments. Hence, it results questionable the existence of an unique model that could properly describe the variety of cuprate superconductors, as the $t$-$t^{\prime }$-$U$ model was thought to be.
\end{abstract}
]

\narrowtext

\section{Introduction}

In recent years much attention has been paid to the physics of electronic
systems with unconventional metallic properties. It is generally believed
that the origin of this anomalous metallic behavior is due to strong
electron correlations in narrow conduction bands\cite{Anderson:1987}. In
this line of thinking many analytical methods have been developed for the
study of strongly correlated electron systems\cite{Fulde}. A parallel
approach to the study of these systems is based on the use of numerical
methods\cite{Dagotto:1994}. The numerical techniques are now very well
developed and many results on finite-dimension lattices are available; these
results are certainly a guide for the construction of a microscopical theory
and to them in any case the different theoretical formulations must refer.

In the last years we have been developing a method of calculation,
denominated Composite Operator Method $\left( \text{{\em COM}}\right) $\cite
{pd,Hubrep,Hubatt,tJ,singlet}, which has been revealed to be a powerful tool
for the description of local and itinerant excitations in strongly
correlated electronic systems. The method is based on the observation that
the original field operators, in terms of which the interacting Hamiltonians
are expressed, are not a convenient basis. Then, a crucial point is the
identification of a set of composite operators that could describe the
quasi-stable excitations which are supposed to be present in the system. The
choice of a non-standard operator basis generates some parameters not
directly connected to the single particle Green's function. Unlikely other
approaches, the presence of these parameters is not inconvenient because it
opens the possibility to bind the dynamics in a suitable Hilbert space,
reabsorbing the symmetries which might be lost when some approximations are
made. Definitely, even if we realize some choice of diagrams, we are capable
to integrate the physical contribution of {\em lost} diagrams by means of
constraining equations. In particular, by using relations with the content
of the Pauli principle\cite{Hubrep}, we are allowed to fix the dynamics of
the system in a fully self-consistent way without recurring to factorization
or other procedures\cite{Roth:1969}. In a physics dominated by the interplay
between the charge and the magnetic configurations, we think that the Pauli
principle should play an important role. Furthermore, the recovery of the
Pauli principle, usually violated by other approximations, assures us that
the hole-particle symmetry is satisfied and that the dynamics is bound to
the right Hilbert space. In fact, the symmetry dictated by the Pauli
principle and that coming from the hole-particle symmetry are intimately
connected, so that the violation of the former implies the violation of the
latter, and viceversa. The method has been applied to the study of several
models: the two-dimensional $p$-$d$ model\cite{pd}, the two-dimensional
repulsive Hubbard model\cite{Hubrep}, the two-dimensional attractive Hubbard
model\cite{Hubatt}, the $t$-$J$ model\cite{tJ} and the two-band singlet-hole
model\cite{singlet}. In this paper we apply the method to the study of the $%
t $-$t^{\prime }$-$U$ model.

Since the discovery of high-$T_c$ superconductivity, there has been a great
deal of discussion about the choice of an effective model suitable to
describe the properties of the copper-oxide planes in the perovskite
structure. Extensive studies of the magnetic properties, showing one spin
degree of freedom in the $Cu$-$O$ plane\cite{Monien:1991}, have resulted in
considerable evidence that the high-temperature superconductors may be
modelled by an effective single-band model. According to this, one of the
most studied model is the single-band Hubbard model which indeed can
qualitatively describe many physical properties experimentally observed in
copper-oxide compounds.

The addition of a finite $t^{\prime }$ diagonal hopping term, that appears
to be material dependent for high-$T_c$ cuprate superconductors, has often
been suggested to handle the complexity of the experimental situation for
the cuprates\cite{Duffy:1995,DagMor}. Moreover, an electron-hole asymmetry
in the next-nearest-neighbor hopping term, combined with a perfect symmetry
of all the other effective parameters, emerges from various reduction
procedures of multi-component electronic models and seems to distinguish the
cuprates from a general charge-transfer insulator\cite
{Feiner:1996,Feiner:1996a}. Recently, it has been argued that this asymmetry
is responsible for the stabilization $\left( \text{destabilization}\right) $
of antiferromagnetic order for electron doping $\left( \text{hole doping}%
\right) $\cite{Tohyama:1994}, whereas the spatial distribution of the doped
carriers\cite{Gooding:1994a} and the damping of quasi particles\cite
{Bala:1995} have been shown to be very sensitive to the sign of $t^{\prime }$%
. A finite $t^{\prime }$ has been found to be essential in reproducing
various experiments $($magnetic structure factor\cite{Benard:1993a,Si:1993},
flat quasi particle dispersion and shape of the Fermi surface\cite
{Dessau:1993} which in turn are responsible for various anomalous normal
state properties, sign change in the Hall effect\cite{Dagotto:1994a},
photoemission data\cite{Uchida:1993}, the behavior of the resistivity with
temperature\cite{Uchida:1993}, the symmetry of the pairing state\cite
{Dagotto:1994}, the actual value of the critical temperature for the optimal
doping concentration\cite{Feiner:1996,Feiner:1996a}$)$. In addition, the
sign of $t^{\prime }$ seems to be relevant for the thermodynamics, in
agreement with more general arguments\cite{Lee:1989} that the propagation
within one sublattice without spin flip allowed by a non zero $t^{\prime }$
would significantly change the physics. Therefore, the next-nearest-neighbor
hopping parameter $t^{\prime }$ emerges as the single parameter, which
carries, at the level of the single band description, the information about
crystal structure outside the $Cu$-$O$ planes and thus differentiates
between the various cuprates\cite{Feiner:1996,Feiner:1996a}.

According to this, we have studied the two-dimensional $t$-$t^{\prime }$-$U$
model to analyze if it could properly describe the variety of cuprate
superconductors.

The plan of the paper is as follows. In Sec.~\ref{model} we present the
two-dimensional $t$-$t^{\prime }$-$U$ model and its properties. Within the
framework of the {\em COM}, in Sec.~\ref{field} we choose a suitable basic
composite field and in Sec.~\ref{Greenfun} we derive the expression of the
propagator for the chosen field. In Sec.~\ref{properties} we present the
comparison of our analytical results for some local properties with
numerical ones obtained by means of the quantum Monte Carlo method on a
finite size lattice\cite{Duffy:1995}. In Sec.~\ref{bands} we show the
results obtained for the structure of the energy bands, the shape of the
Fermi surface and the relative position of the van Hove singularity with
respect to the Fermi level as functions of the model parameters. We have
payed particular attention to the comparison with the experimental data
available for the superconducting cuprates. In Sec.~\ref{conclusions} some
conclusions are given.

\section{The model}

\label{model}

The two-dimensional $t$-$t^{\prime }$-$U$ model is described by the
following Hamiltonian: 
\begin{equation}
H=\sum_{ij}t_{ij}c^{\dagger }\left( i\right) c\left( j\right)
+U\sum_in_{\uparrow }\left( i\right) n_{\downarrow }\left( i\right) -\mu
\sum_in\left( i\right)  \label{hamiltonian}
\end{equation}
where $c^{\dagger }\left( i\right) =\left( c_{\uparrow }^{\dagger }\left(
i\right) ,c_{\downarrow }^{\dagger }\left( i\right) \right) $ is the
electron operator on the site $i$ in the spinor notation , $n_\sigma \left(
i\right) $ is the charge-density operator for the spin $\sigma $ and $%
n\left( i\right) $ is the total charge-density operator. In the hopping
matrix $t_{ij}$, the terms up to the next-nearest neighbors, situated along
the plaquette diagonals, have to be retained. The $U$ parameter represents
the intrasite Coulomb potential and $\mu $ the chemical potential. This
model does not enjoy the hole-particle symmetry owing to the presence of the 
$t^{\prime }$ term, that, under the hole-particle transformation $\left(
c^{\dagger }\left( i\right) \rightarrow \left( -1\right) ^ic\left( i\right)
\right) $, changes its sign: 
\begin{equation}
\mu \left( n,t^{\prime }\right) =U-\mu \left( 2-n,-t^{\prime }\right) \text{.%
}  \label{hole}
\end{equation}

\section{The basic field}

\label{field}

Let us introduce the following basic field: 
\begin{equation}
\psi \left( i\right) =\left( 
\begin{array}{l}
\xi \left( i\right) \\ 
\eta \left( i\right)
\end{array}
\right) =\left( 
\begin{array}{c}
\left( 1-n\left( i\right) \right) c\left( i\right) \\ 
n\left( i\right) c\left( i\right)
\end{array}
\right)  \label{basic}
\end{equation}
where the composite electron operators $\xi \left( i\right) $ and $\eta
\left( i\right) $ represent the $n\left( i\right) =0\leftrightarrow n\left(
i\right) =1$ and the $n\left( i\right) =1\leftrightarrow n\left( i\right) =2$
restricted electronic transitions, respectively. They make up the {\em %
so-called} Hubbard operator doublet $\left( c\left( i\right) =\xi \left(
i\right) +\eta \left( i\right) \right) $. These two composite electron
operators are well recognized to be responsible for the main distribution of
the electron density of states for the Hubbard model from both analytical
and numerical calculations\cite{Dagotto:1994}.

The field $\psi $ satisfies the following equation of motion obtained from
the Hamiltonian $\left( \ref{hamiltonian}\right) $: 
\begin{equation}
\ i\frac \partial {\partial t}\psi \left( i\right) =\left( 
\begin{array}{c}
-\mu \xi \left( i\right) +\sum_jt_{ij}c\left( j\right) +\pi \left( i\right)
\\ 
-\left( \mu -U\right) \eta \left( i\right) -\pi \left( i\right)
\end{array}
\right)
\end{equation}
where the operator $\pi $ has the following form 
\begin{equation}
\pi \left( i\right) =\sum_jt_{ij}\left( \frac 12\sigma ^\mu n_\mu \left(
i\right) c\left( j\right) +c\left( i\right) \left( c^{\dagger }\left(
j\right) c\left( i\right) \right) \right) \text{.}  \label{pai}
\end{equation}
The following definitions have been used 
\begin{equation}
\sigma _\mu =\left( {\bf 1},{\bf \sigma }\right) \qquad n_\mu \left(
i\right) =c^{\dagger }\left( i\right) \sigma _\mu c\left( i\right)
\end{equation}
with ${\bf 1}$ and ${\bf \sigma }$ being the unity and the three Pauli
matrices respectively and $n_\mu \left( i\right) $ representing for $\mu =0$
the total charge- and for $\mu =1,2,3$ the spin- density operator for the
site $i$. In Eq.~\ref{pai} and in ones that will follow the summation with
respect to greek indices is understood.

\section{The Green's function}

\label{Greenfun}

The properties of the system are conveniently expressed in terms of the
single particle retarded thermal Green's function: 
\begin{equation}
S\left( {\bf k},\omega \right) =\left\langle R\left[ \psi \left( i\right)
\psi ^{\dagger }\left( j\right) \right] \right\rangle _{F.T.}
\end{equation}
where $\left\langle \quad \right\rangle _{F.T.}$ is the Fourier transform of
the thermal average and $R\left[ \quad \right] $ indicates the retarded
time-ordered product. In the framework of the {\em COM} and neglecting
finite life-time effects\cite{Hubrep} we have 
\begin{equation}
S\left( {\bf k},\omega \right) =\frac 1{\omega -m\left( {\bf k}\right)
I\left( {\bf k}\right) ^{-1}}I\left( {\bf k}\right) \text{.}  \label{Greenk}
\end{equation}
By considering paramagnetic contour conditions together with the
roto-translational invariance, $I\left( {\bf k}\right) $ has the explicit
expression: 
\begin{equation}
I\left( {\bf k}\right) =\left\langle \left\{ \psi \left( i\right) ,\psi
^{\dagger }\left( j\right) \right\} \right\rangle _{F.T.}=\left( 
\begin{array}{cc}
\left( 1-\frac n2\right) & {\bf 0} \\ 
{\bf 0} & \frac n2
\end{array}
\right)
\end{equation}
$n$ being the thermal average of the total charge-density operator $\left(
n=\left\langle n\left( i\right) \right\rangle \right) $. Moreover, $m\left( 
{\bf k}\right) $ is defined as 
\begin{equation}
m\left( {\bf k}\right) =\left\langle \left\{ i\frac \partial {\partial t}%
\psi \left( i\right) ,\psi ^{\dagger }\left( j\right) \right\} \right\rangle
_{F.T.}\text{.}  \label{mmatrix}
\end{equation}
From Eq.~\ref{mmatrix} direct calculations give \FL
\begin{mathletters}
\begin{eqnarray}
m_{11}\left( {\bf k}\right) &=&-\mu \left( 1-\frac n2\right) -4t\left(
\Delta +\alpha \left( {\bf k}\right) \left( 1-n+p\right) \right)  \nonumber
\\
&&\ \ \ -4t^{\prime }\left( \Delta ^{\prime }+\beta \left( {\bf k}\right)
\left( 1-n+p^{\prime }\right) \right) \\
m_{12}\left( {\bf k}\right) &=&m_{21}\left( {\bf k}\right) =4t\left( \Delta
-\alpha \left( {\bf k}\right) \left( \frac n2-p\right) \right) {\bf +} 
\nonumber \\
&&\ \ \ {\bf +}4t^{\prime }\left( \Delta ^{\prime }-\beta \left( {\bf k}%
\right) \left( \frac n2-p^{\prime }\right) \right) \\
m_{22}\left( {\bf k}\right) &=&-\left( \mu -U\right) \frac n2-4t\left(
\Delta +\alpha \left( {\bf k}\right) p\right) {\bf +}  \nonumber \\
&&\ \ \ -4t^{\prime }\left( \Delta ^{\prime }+\beta \left( {\bf k}\right)
p^{\prime }\right)
\end{eqnarray}
with \FL
\end{mathletters}
\begin{mathletters}
\begin{eqnarray}
\Delta &=&\left\langle \xi _{\uparrow }^\alpha \left( i\right) \xi
_{\uparrow }^{\dagger }\left( i\right) \right\rangle -\left\langle \eta
_{\uparrow }^\alpha \left( i\right) \eta _{\uparrow }^{\dagger }\left(
i\right) \right\rangle \\
\Delta ^{\prime } &=&\left\langle \xi _{\uparrow }^\beta \left( i\right) \xi
_{\uparrow }^{\dagger }\left( i\right) \right\rangle -\left\langle \eta
_{\uparrow }^\beta \left( i\right) \eta _{\uparrow }^{\dagger }\left(
i\right) \right\rangle \\
p &=&\frac 14\left\langle n_\mu ^\alpha \left( i\right) n_\mu \left(
i\right) \right\rangle  \nonumber \\
&&-\left\langle \left( c_{\uparrow }\left( i\right) c_{\downarrow }\left(
i\right) \right) ^\alpha c_{\downarrow }^{\dagger }\left( i\right)
c_{\uparrow }^{\dagger }\left( i\right) \right\rangle \\
p^{\prime } &=&\frac 14\left\langle n_\mu ^\beta \left( i\right) n_\mu
\left( i\right) \right\rangle  \nonumber \\
&&-\left\langle \left( c_{\uparrow }\left( i\right) c_{\downarrow }\left(
i\right) \right) ^\beta c_{\downarrow }^{\dagger }\left( i\right)
c_{\uparrow }^{\dagger }\left( i\right) \right\rangle \text{.}
\end{eqnarray}
We are using the following notation 
\end{mathletters}
\begin{equation}
\zeta ^\alpha \left( i\right) =\sum_j\alpha _{ij}\zeta \left( j\right) \quad
\zeta ^\beta \left( i\right) =\sum_j\beta _{ij}\zeta \left( j\right) \text{,}
\end{equation}
$\alpha \left( {\bf k}\right) $ and $\beta \left( {\bf k}\right) $ are the
Fourier transforms of $\alpha _{ij}$ and $\beta _{ij}$. These latter
represent the projectors on the nearest and next-nearest neighbors,
respectively.

The {\em internal} parameters $\Delta $, $\Delta ^{\prime }$, $p$ and $%
p^{\prime }$, as the chemical potential $\mu $, have to be calculated in
order to obtain the fermionic propagator as a function of the {\em external}
parameters $t$, $t^{\prime }$, $U$, $n$ and $T$ (temperature).

The chemical potential $\mu $ can be calculated using the equation that
gives the filling $n$ 
\begin{equation}
n=2\left( 1-\left\langle \xi _{\uparrow }\left( i\right) \xi _{\uparrow
}^{\dagger }\left( i\right) \right\rangle -\left\langle \eta _{\uparrow
}\left( i\right) \eta _{\uparrow }^{\dagger }\left( i\right) \right\rangle
\right) \text{.}
\end{equation}
The parameters $\Delta $ and $\Delta ^{\prime }$ can be determined by their
definitions that give a direct connection with the elements of the Green's
function. The parameters $p$ and $p^{\prime }$ are not directly calculable
starting from the elements of the single particle Green's function and they
will be calculated using a relation with the content of the Pauli principle%
\cite{Hubrep} 
\begin{equation}
\left\langle \xi \left( i\right) \eta ^{\dagger }\left( i\right)
\right\rangle ={\bf 0}\text{.}  \label{Pauli}
\end{equation}

The use of the symmetry contained in Eq.~\ref{Pauli} allows us to fix the
dynamics of the system in a fully self-consistent way.

\section{The local properties}

\label{properties}

\subsection{The chemical potential}

\begin{figure}[tb]
\centerline{\psfig{figure=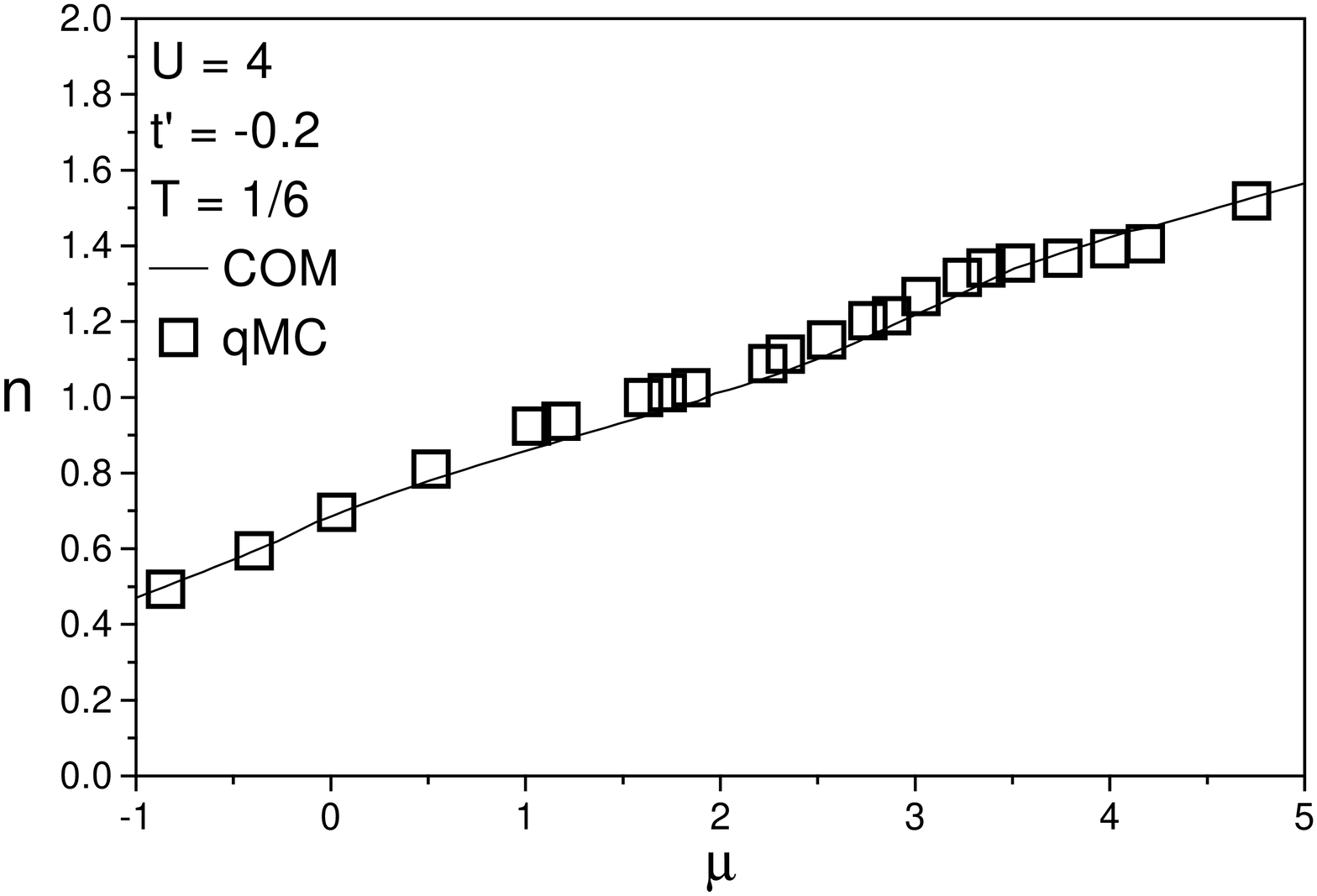,width=8cm,angle=270,clip=}}
\caption{Chemical potential $\mu $ as a function of the filling $n$ for $U=4$ , $T=\frac 16$ and $t^{\prime }=-0.2 $. COM indicates our results, 
qMC the ones of Ref.~\protect\onlinecite{Duffy:1995}}
\label{mormu}
\end{figure}

We have calculated the chemical potential $\mu $ as a function of the
external parameters using the system of self-consistent equations described
in Sec.~\ref{Greenfun}. We have compared our results with the ones obtained
by means of the quantum Monte Carlo method\cite{Duffy:1995} on a finite size
lattice $8\times 8$, see Fig.~\ref{mormu}.

The quantum Monte Carlo data present a plateau for $n\approx 1.3$. This
plateau is completely absent in our curve and is related to a finite size
effect\cite{Duffy}. Moreover, the opening of the antiferromagnetic gap, due
to a spin density wave instability\cite{Duffy}, does not allow us to
reproduce the behavior near half-filling where our solution is paramagnetic;
this latter comment will also reflect on the comparison done with the double
occupancy data.

Finally, it is interesting to point out that our theoretical calculations
predict a change, driven by the value of $t^{\prime }$, of less than $1\%$
in the critical value of $U$ which signs the opening of the Mott-Hubbard gap.

\subsection{The double occupancy}

The double occupancy $D$, defined as the probability to have a couple of
electrons $\left( \uparrow \downarrow \right) $ on the same site, can be
calculated through the following equation 
\begin{eqnarray}
D=\left\langle n_{\uparrow }\left( i\right) n_{\downarrow }\left( i\right)
\right\rangle =\frac 12\left\langle \eta ^{\dagger }\left( i\right) \eta
\left( i\right) \right\rangle \text{.}
\end{eqnarray}

\begin{figure}[tb]
\centerline{\psfig{figure=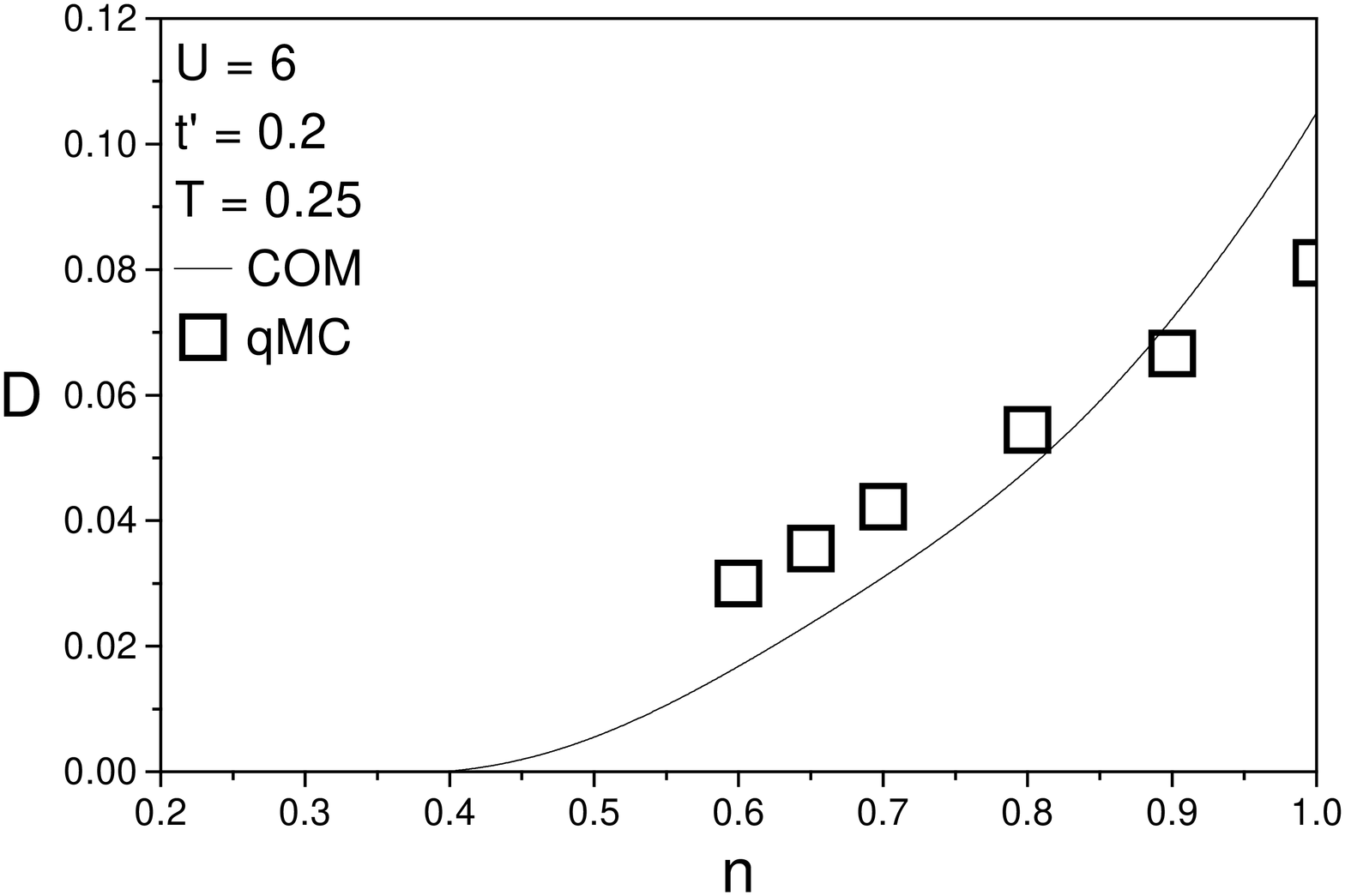,width=8cm,angle=270,clip=}}
\caption{ Double occupancy $D$ as a function of the filling $n$ for $U=6t$, $ T=\frac 14$ and $t^{\prime }=0.2$. COM indicates our results, 
qMC the ones obtained in Ref.~\protect\onlinecite{Duffy:1995}}
\label{mordo}
\end{figure}

We have compared our results with the ones obtained by means of the quantum
Monte Carlo method\cite{Duffy:1995} on a finite size lattice $8\times 8$,
see Fig.~\ref{mordo}. In the low filling region our solution presents a
characteristic feature, the existence of a critical value of the filling $%
n_D\left( U\right) $ before which the double occupancy is almost zero $%
\left( n\leq n_D\left( U\right) \right) $. The small residual is due to the
thermal fluctuations. This kind of behavior seems to be absent in the
quantum Monte Carlo data at finite temperature, but it can be surely
inferred when the zero temperature quantum Monte Carlo data for the chemical
potential $\mu $ are reported as a function of the $U$ parameter\cite{Hubrep}%
.

\begin{figure}[tb]
\centerline{\psfig{figure=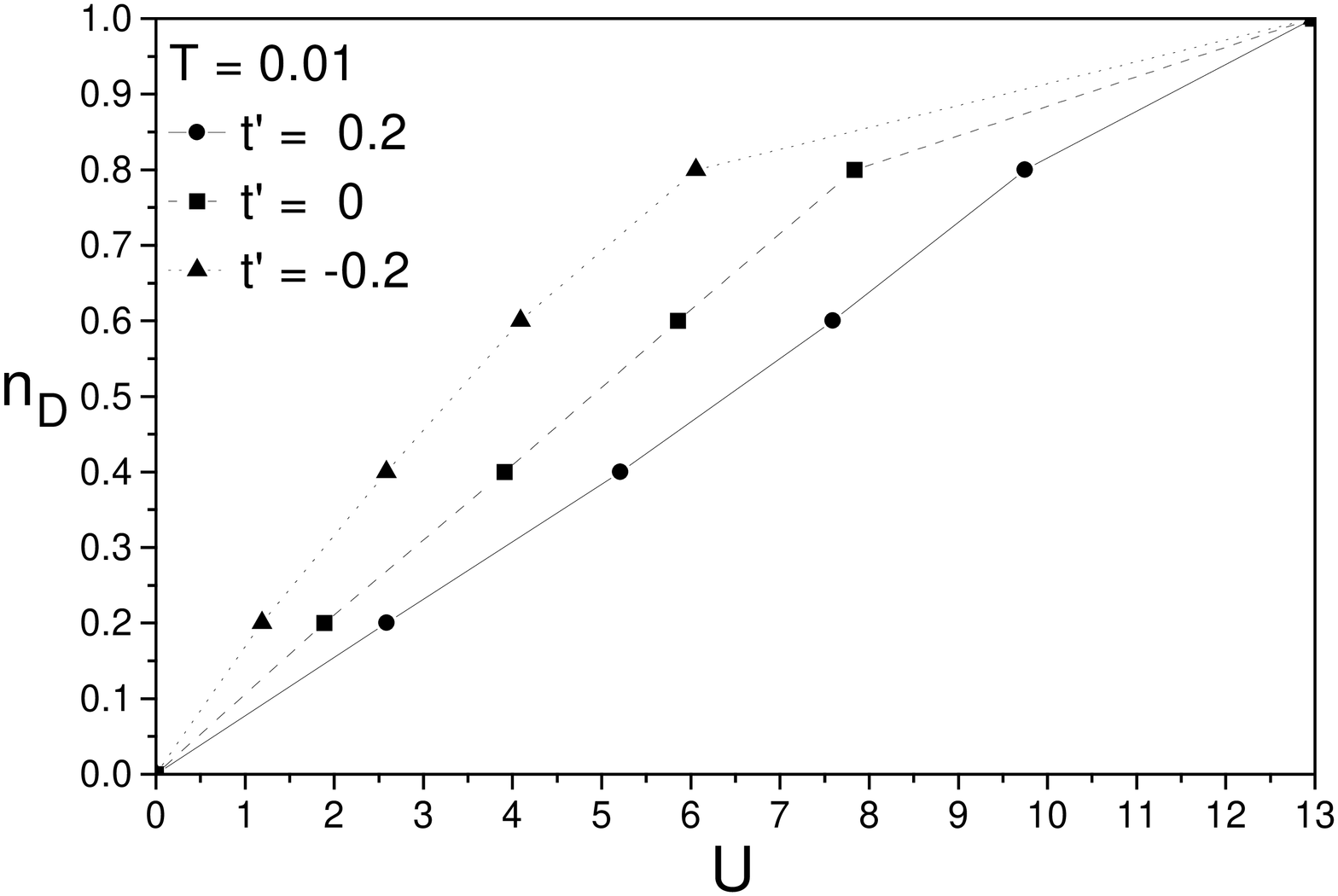,width=8cm,angle=270,clip=}}
\caption{Critical value of the filling $n_D$ as a
function of the intrasite Coulomb potential $U$ for $T=0.01$ and $t^{\prime }=-0.2,0,0.2$.}
\label{xdU}
\end{figure}

The explanation of this behavior of the double occupancy can be given as the
existence of two regimes: one in which the low filling let the carriers the
possibility to move freely and avoid the high-energetic double occupancy of
some sites and another one in which the number of carriers excludes the
possibility to avoid the double occupancy. Obviously, this value of the
filling, which marks the changing of regime, depends strongly on the value
of the $U$ and $t^{\prime }$ parameters, see Fig.~\ref{xdU}.

\subsection{The energy per site}

The total energy per site $E_s$ can be written as 
\begin{equation}
E_s=T_s+V_s
\end{equation}
where 
\begin{mathletters}
\begin{eqnarray}
T_s &=&\frac 1N\sum_\sigma \sum_{ij}t_{ij}\left\langle c_\sigma ^{\dagger
}\left( i\right) c_\sigma \left( j\right) \right\rangle \\
V_s &=&U\frac 1N\sum_i\left\langle n_{\uparrow }\left( i\right)
n_{\downarrow }\left( i\right) \right\rangle =UD
\end{eqnarray}
are the kinetic and the potential energies, respectively.

\begin{figure}[tb]
\centerline{\psfig{figure=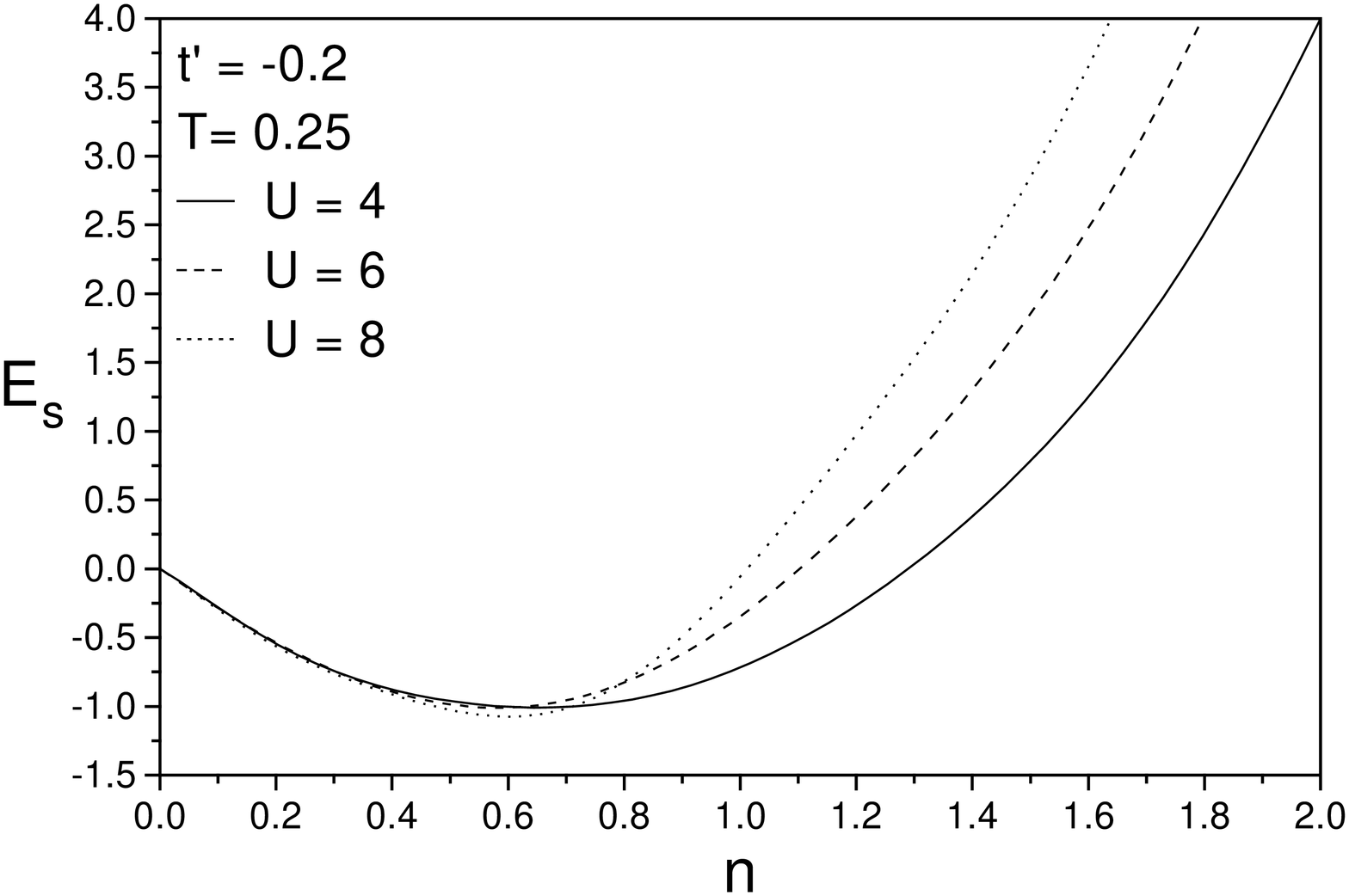,width=8cm,angle=270,clip=}}
\caption{Energy per site $E_s$ as a function of the filling $n$ for $ t^{\prime }=-0.2$, $T=0.25$ and $U=4,6,8$.}
\label{ok}
\end{figure}

The behavior of the total energy per site $E_s$ follows that of the kinetic
energy $T_s$ for low values of the filling, where the double occupancy is
almost zero whereas it follows that of the potential energy $V_s$ for values
of the filling greater than $n_D\left( U\right) $, see Fig.~\ref{ok}. This
kind of behavior has been also found by means of quantum Monte Carlo method
calculations\cite{Moreo:1990,Dagotto:1992a} and slave-boson ones\cite
{Lilly:1990} giving a further confirmation about the presence of a region of
the filling where the double occupancy has almost zero value.

\section{The band properties}

\label{bands}

In the framework of the {\em COM}, the Fourier transform of the single
particle retarded thermal Green's function, see Eq.~\ref{Greenk}, may be
rewritten as: 
\end{mathletters}
\begin{equation}
S\left( {\bf k},\omega \right) =\sum_{i=1}^2\frac{\sigma ^i\left( {\bf k}%
\right) }{\omega -E_i\left( {\bf k}\right) }
\end{equation}
where $\sigma ^{1,2}\left( {\bf k}\right) $ are given by 
\begin{mathletters}
\begin{eqnarray}
\sigma _{11}^i\left( {\bf k}\right) &=&\frac{I_{11}\left( 2Q\left( {\bf k}%
\right) +\left( -\right) ^{i+1}\Delta \Sigma \left( {\bf k}\right) \right) }{%
4Q\left( {\bf k}\right) } \\
\sigma _{12}^i\left( {\bf k}\right) &=&\sigma _{21}^i\left( {\bf k}\right)
=\left( -\right) ^{i+1}\frac{m_{12}\left( {\bf k}\right) }{2Q\left( {\bf k}%
\right) } \\
\sigma _{22}^i\left( {\bf k}\right) &=&\frac{I_{22}\left( 2Q\left( {\bf k}%
\right) +\left( -\right) ^i\Delta \Sigma \left( {\bf k}\right) \right) }{%
4Q\left( {\bf k}\right) }
\end{eqnarray}
with

\end{mathletters}
\begin{mathletters}
\begin{eqnarray}
Q\left( {\bf k}\right) &=&\frac 12\sqrt{\left( U-\frac{m_{12}\left( {\bf k}%
\right) }{I_{11}I_{22}}\right) ^2+2nU\frac{m_{12}\left( {\bf k}\right) }{%
I_{11}I_{22}}} \\
R\left( {\bf k}\right) &=&\frac 12\left( U-2\mu -8t\alpha \left( {\bf k}%
\right) -8t^{\prime }\beta \left( {\bf k}\right) \right)  \nonumber \\
&&-\frac 12\frac{m_{12}\left( {\bf k}\right) }{I_{11}I_{22}} \\
\Delta \Sigma \left( {\bf k}\right) &=&\left( 1-n\right) \frac{m_{12}\left( 
{\bf k}\right) }{I_{11}I_{22}}-U\text{.}
\end{eqnarray}
$E_{1,2}\left( {\bf k}\right) $ represent the energy bands and have the
following expressions: 
\end{mathletters}
\begin{equation}
E_i\left( {\bf k}\right) =R\left( {\bf k}\right) +\left( -\right)
^{i+1}Q\left( {\bf k}\right) \text{.}
\end{equation}

According to this, the Fermi surface of the system may be defined as $%
E_2\left( {\bf k}\right) =0$ and the electronic density of states, $N\left(
\omega \right) $, may be computed through the following formula: 
\begin{equation}
N\left( \omega \right) =\frac 1{2\pi ^2}\int d{\bf k~}\left( {\bf -}\frac 1%
\pi \sum_{i,j=1}^2\mathop{\rm Im}S_{ij}\left( {\bf k},\omega \right) \right)  \label{dens}
\end{equation}
where the integration has to be performed on the Brillouin zone. The
presence of two bands gives a structure of the density of states
characterized by two logarithmic van Hove singularities.

\subsection{The energy bands and the van Hove singularity}

Both band structure calculations and experiments generally find that the
Fermi level is close to the van Hove singularity at the optimal doping $%
\left( \delta _c\right) $ for the majority of the multi-layer cuprate
superconductors, like $YBa_2Cu_3O_{7-\delta }$ $\left( YBCO\right) $, $%
Bi_2Sr_2Ca_1Cu_2O_\delta $ $\left( Bi\text{-}2212\right) $ and $Hg$ compounds%
\cite{Markiewicz,Dessau:1993,Abrikosov:1993}.

In particular, for $YBCO$ the relative distance $\left( \Delta E\right) $
between the Fermi level and van Hove singularity has been found to be within 
$6$~$meV$ for its optimal doping concentration\cite{Gofron:1994} $\left(
\delta _c^{YBCO}\approx 0.15\right) $.

For the electron-doped $Nd_{2-x}Ce_xCuO_4$ $\left( NCCO\right) $, it has
been found a value of $\Delta E$ of $\approx 200~meV$ for its optimal
electron-doping concentration\cite{King:1993} $\left( x_c^{NCCO}\approx
0.15\right) $.

So far, no photoemission results are available for the lanthanates due to
the difficulty in obtaining high enough quality samples. However, for the $%
La_{2-x}Sr_xCuO_4$ $\left( LSCO\right) $ the van Hove singularity seems to
coincide with the Fermi level at the critical doping $\left( x_c\right) $ at
which the superconductivity disappears $\left( x_c^{LSCO}\approx 0.3\right) $
after the experimental data for the static susceptibility\cite{Torrance:1989}%
, the electronic specific heat and the entropy\cite{Loram:1996}.

More generally, the experimentally derived dispersions of the $Cu$-$O$ plane
anti-bonding bands for a series of cuprates ($Bi$-$2212$, $YBCO$, $%
Bi_2Sr_2CuO_\delta $ $\left( Bi\text{-}2201\right) $, $YBa_2Cu_4O_8$, $NCCO$%
) show a remarkable similarity to one another with the van Hove singularity
appearing near the $Y$-point $\left( 0,\pi \right) $ of the Brillouin zone%
\cite{Shen:95}.

\begin{figure}[tb]
\centerline{\psfig{figure=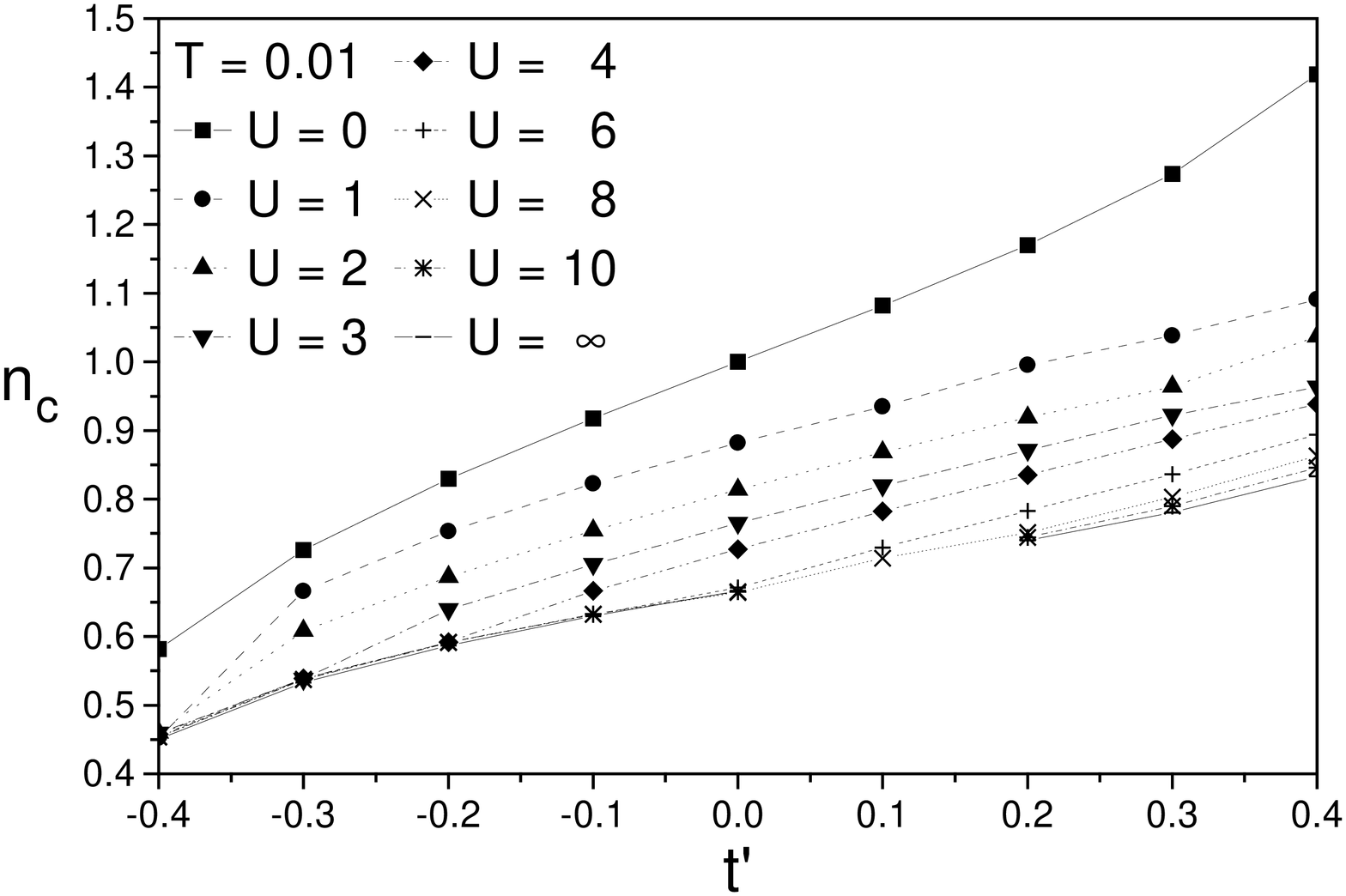,width=8cm,angle=270,clip=}}
\caption{Critical filling $n_c$ as a function of the $t^{\prime }$ and 
$U$ parameters for $T=0.01$.}
\label{nc}
\end{figure}

The value of the $t^{\prime }$ parameter is crucial in determining the
structure of the energy bands and therefore the relative position of the van
Hove singularity with respect to the Fermi level. The critical value of the
filling $\left( n_c\right) $ for which the van Hove singularity of the lower
band coincides with the Fermi level, has the behavior shown in Fig.~\ref{nc}
as a function of the $t^{\prime }$ and $U$ parameters. $n_c$ has been
computed by studying the density of states, see Eq.~\ref{dens}, as a
function of the filling $n$ for fixed values of the $t^{\prime }$ and $U$
parameters. Fig.~\ref{nc} contains a complete information about the
structure and the doping evolution of the density of states and therefore
permits a comprehensive comparison between the experimental situation and
the physics described by the $t$-$t^{\prime }$-$U$ model.

The critical value of the filling $\left( n_c^{\prime }\right) $, that
corresponds to the coincidence of the upper band van Hove singularity with
the Fermi level can be obtained by the one of the lower band through the
following formula, that comes directly from the hole-particle symmetry 
\begin{equation}
n_c^{\prime }\left( U,t^{\prime }\right) =n_c^{\prime }\left( 0,t^{\prime
}\right) +\left( n_c\left( 0,-t^{\prime }\right) -n_c\left( U,-t^{\prime
}\right) \right) \text{.}
\end{equation}
This critical value of the filling can be also interesting with respect to
the electron-doped compounds for which the relevant band is the upper one.

\begin{figure}[tb]
\centerline{\psfig{figure=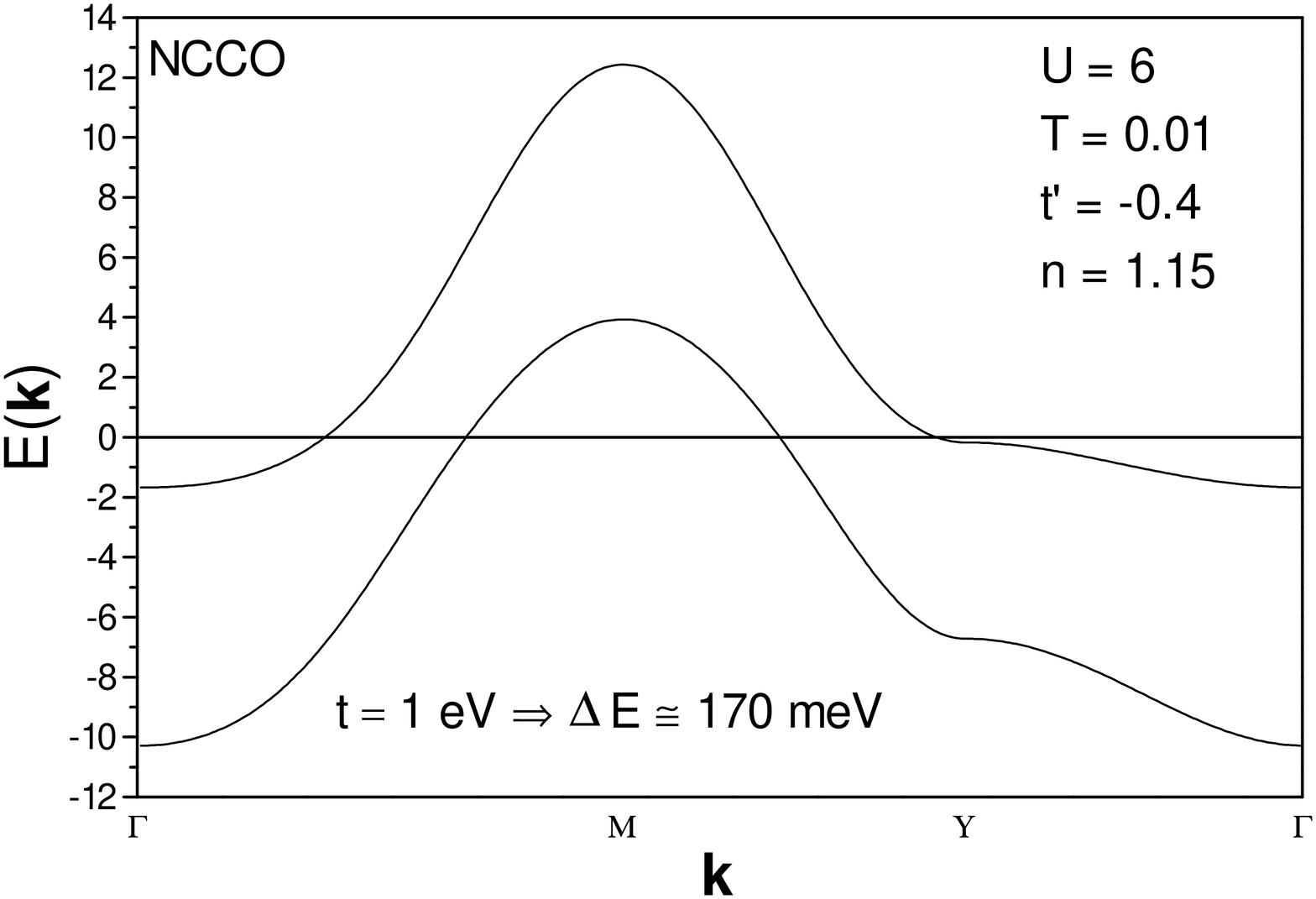,width=8cm,angle=270,clip=}}
\caption{Energy bands for $t^{\prime }=-0.4$, $T=0.01$, $n=1.15$ and $U=6$.}
\label{NCCO1}
\end{figure}

\begin{figure}[tb]
\centerline{\psfig{figure=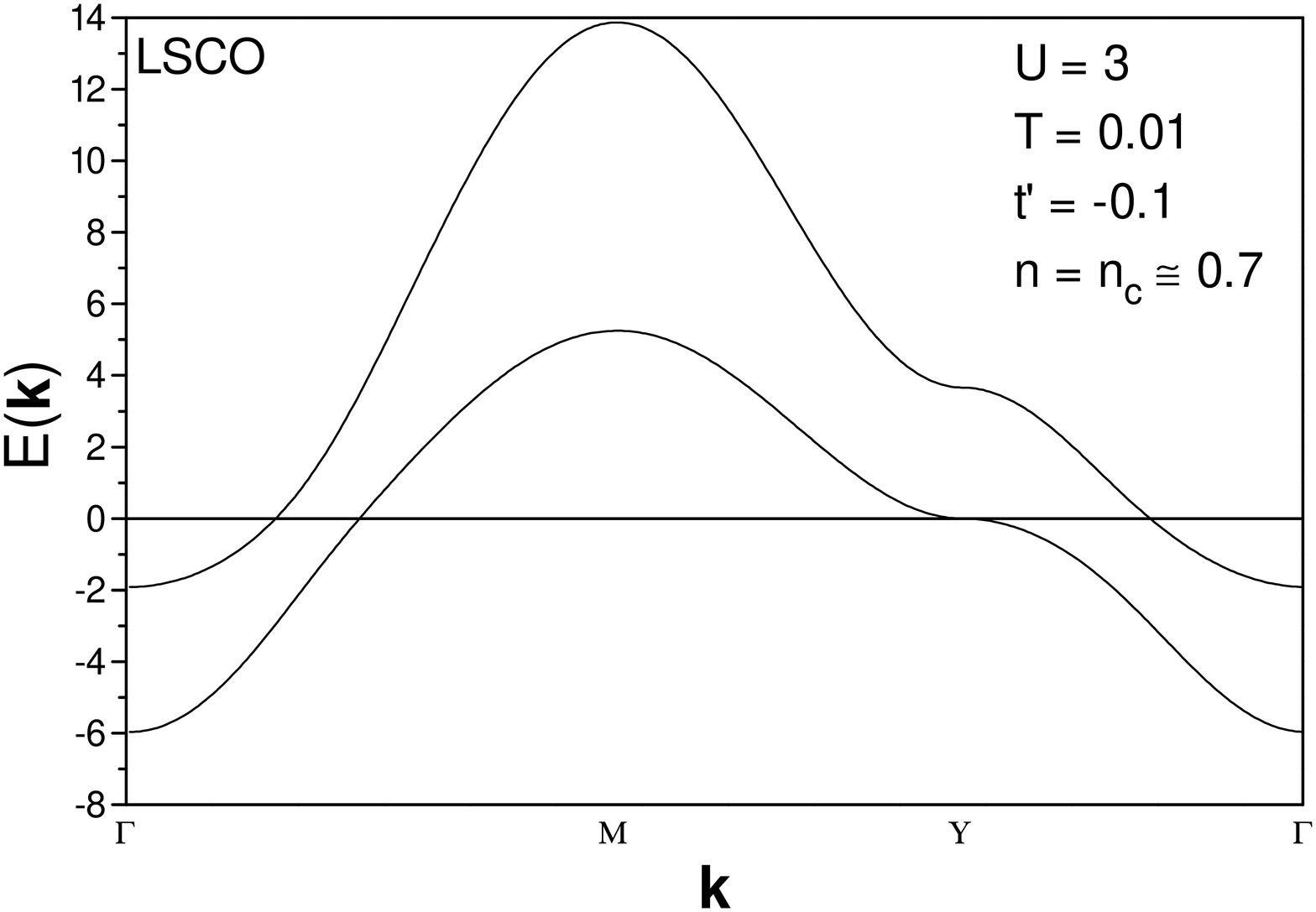,width=8cm,angle=270,clip=}}
\caption{Energy bands for $t^{\prime }=-0.1$, $T=0.01$, $n=0.7$ and $U=3$.}
\label{LSCO1}
\end{figure}

We have studied the structure of the energy bands as function of the model
parameters. The results have shown that is possible to obtain a good
agreement with the experimental data by choosing reasonable sets of
parameters. In particular, the value of $\Delta E$ for $NCCO$ together with
the shape of the energy band can be obtained by the following set of
parameters: $U=6$,\ $t^{\prime }=-0.4$, see Fig.~\ref{NCCO1}. In the case of 
$LSCO$, we can obtain the right value for $x_c$ by using the following set
of parameters: $U=3$, $t^{\prime }=-0.1$, see Fig.~\ref{LSCO1}. In this
figure we can observe the coincidence between the van Hove singularity and
the Fermi level as required by the ${\em LDA}$ calculations\cite{Xu:1987}.

Finally, it can be easily seen from Fig.~\ref{nc}, that it is impossible to
obtain the features suggested for $YBCO$ unless to use a set of parameters
like: $U=1$, $t^{\prime }=-0.4$ with a value of the $U$ parameter really too
small in comparison with the band calculation results\cite
{Feiner:1996,Feiner:1996a}. Moreover, even using this set of parameters, the
relevant van Hove singularity results to be the upper band one, in strict
contradiction with the hole-doped nature of the compound. This is due to the
value of the $t^{\prime }$ parameter necessary to obtain the right bending
of the Fermi surface after the {\em ARPES} data\cite{Liu:1992}. A value of $%
-0.4$ for the $t^{\prime }$ parameter gives a value for the critical filling
of the lower band van Hove singularity too small with respect to the optimal
doping concentration required by experimental data\cite{Liu:1992}.

Independently to the chosen set of parameters the van Hove singularity
appears at the $Y$-point as in the experimental case.

\subsection{The Fermi surface}

The Fermi surface of the various cuprates are remarkably similar one to
another; in particular, photoemission experiments show a large Fermi surface
for a series of cuprates at their optimal doping concentration\cite{Shen:95}
($Bi$-$2212$, $Bi$-$2201$, $NCCO$, $YBCO$).

Photoemission studies of $NCCO$ find a hole-like and roughly circular Fermi
surface\cite{King:1993}. The apparent simplicity of this Fermi surface is
deceptive, since the transport properties imply that the majority carriers
are electron-like.

Positron annihilation studies of the doping dependence of the Fermi surface
for the $LSCO$ are consistent with a pseudo-nested hole-like Fermi surface%
\cite{Howell:1994} predicted by {\em LDA} calculations\cite{Xu:1987}.

\begin{figure}[tb]
\centerline{\psfig{figure=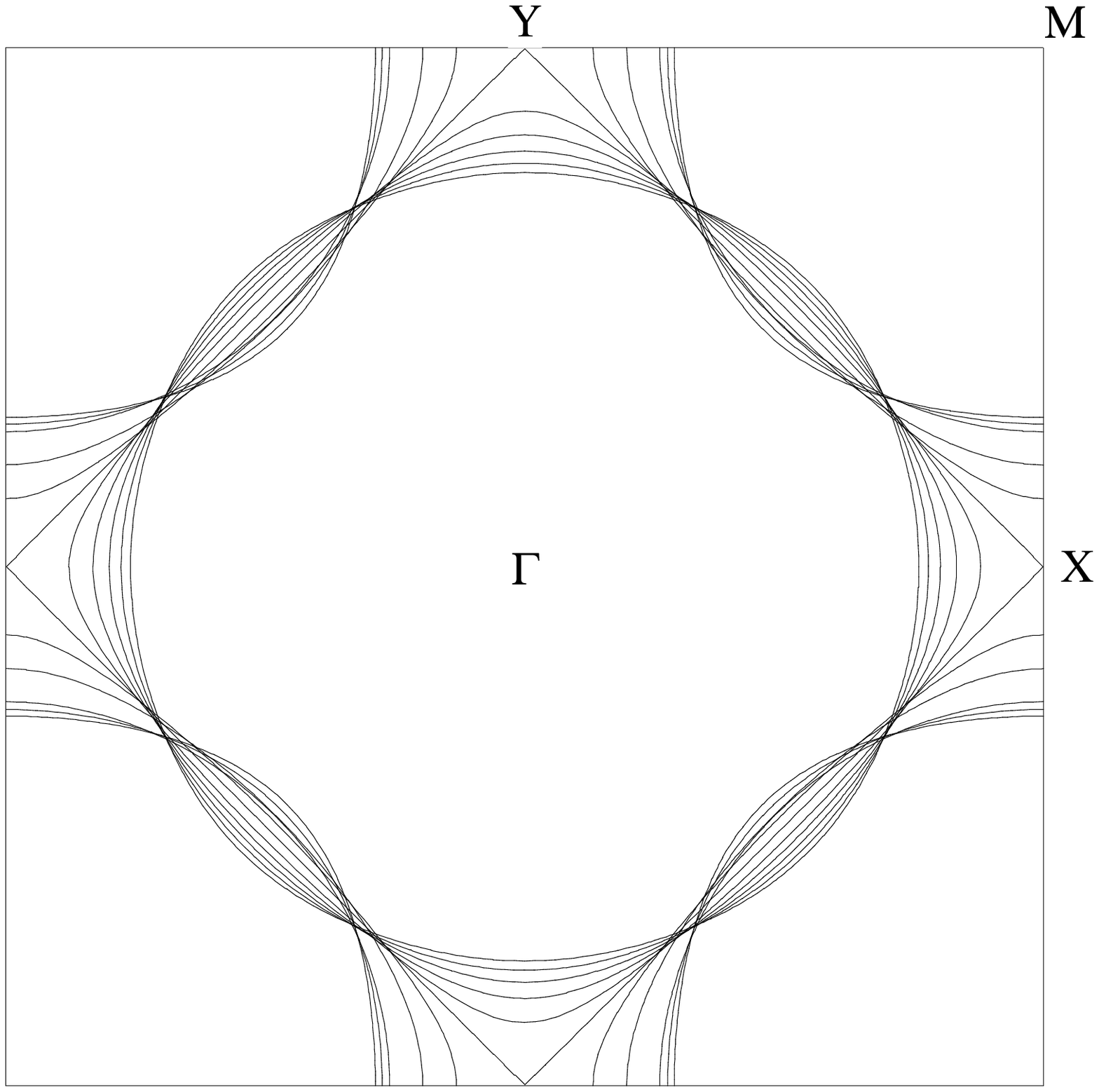,width=8cm,bbllx=16pt,bblly=143pt,bburx=579pt,bbury=706pt,clip=}}
\caption{FS for $t^{\prime }=-0.5 \rightarrow 0.5$, $T=0.01$, $n=0.73$ and $U=4$.}
\label{rotation}
\end{figure}

The shape and, in particular, the bending of the Fermi surface are strongly
dependent on the value of the $t^{\prime }$ parameter. The bending is
electron-like for positive values of $t^{\prime }$ and hole-like for
negative ones, independently on the strength of the $U$ parameter. This can
lead for fixed values of filling $n$ and of the $U$ parameter to a real
rotation of the Fermi surface by varying the value of the $t^{\prime }$
parameter. In Fig.~\ref{rotation} we show the Fermi surface of the $t$-$%
t^{\prime }$-$U$ model with $U=4$, $T=0.01$ and $n\cong 0.73$ for values of
the $t^{\prime }$ parameter that range from $-0.5$ to $0.5$ with step $0.1$.
The chosen value of the filling corresponds to the critical value $n_c$ for $%
t^{\prime }=0$. The Fermi surface is open and hole-like for $t^{\prime
}=-0.5 $. It is nested for $t^{\prime }=0$. It is closed and electron-like
for $t^{\prime }=0.5$. This gives the idea of a $\frac \pi 4$ possible
rotation that can be driven by the $t^{\prime }$ parameter.

The critical value of the filling for which the Fermi surface closes
corresponds to the value for which the van Hove singularity of the lower
band coincides with the Fermi level. The perfect nesting can be obtained
only for a zero value of the $t^{\prime }$ parameter. Any non-zero value
leads to a pseudo-nesting as the Fermi surface, although closed, conserves
some bending.

\begin{figure}[tb]
\centerline{\psfig{figure=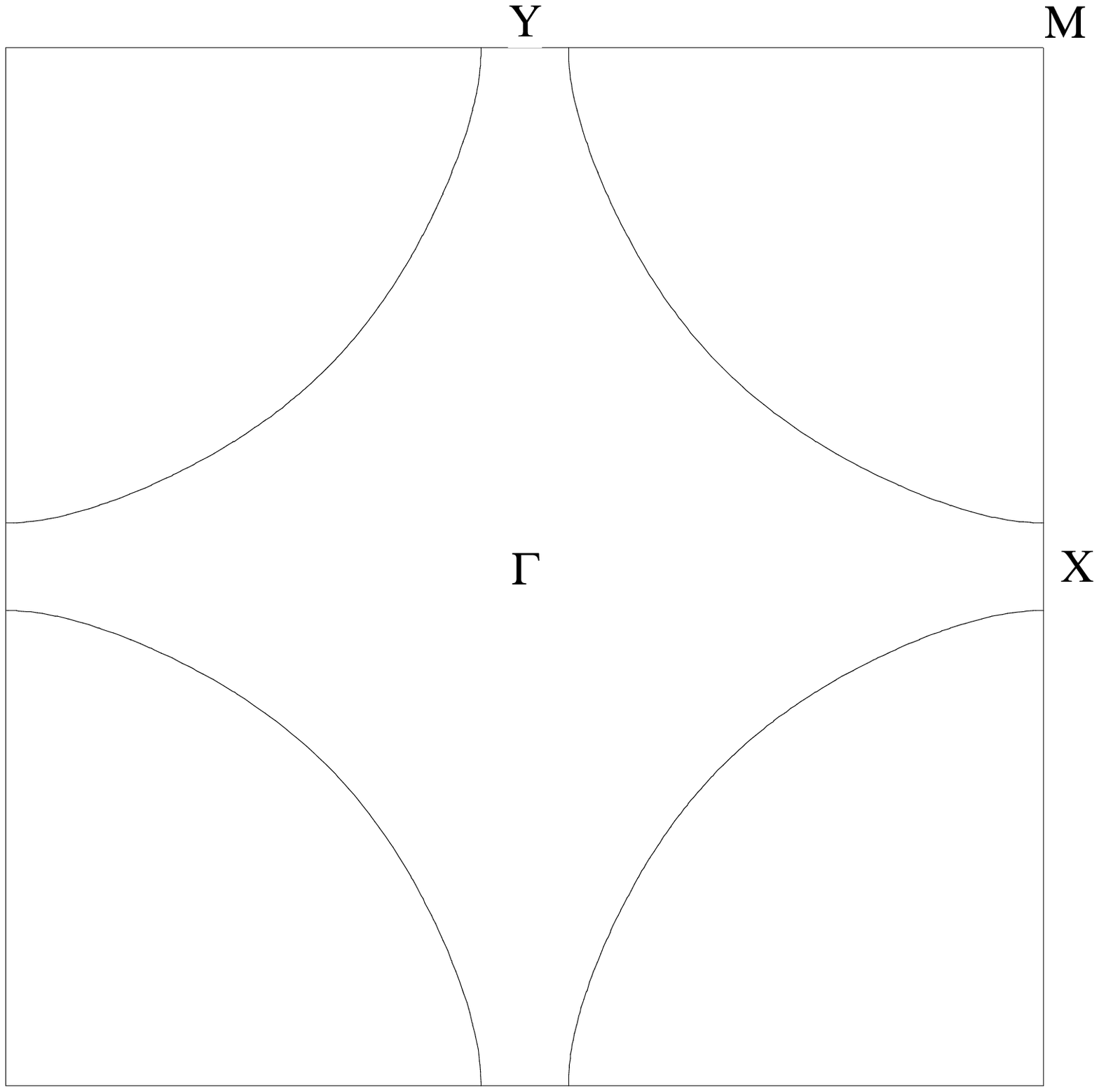,width=8cm,bbllx=16pt,bblly=143pt,bburx=579pt,bbury=706pt,clip=}}
\caption{FS for $t^{\prime }=-0.4$, $T=0.01$, $n=1.15$ and $U=6$.}
\label{NCCO2}
\end{figure}

It is really relevant that the experimentally observed Fermi surface for $%
NCCO$ can be obtained, in our formulation, by the same set of parameters
that gives a correct band dispersion, see Fig.~\ref{NCCO2}. Moreover, it has
to be pointed out that the value of the $t^{\prime }$ parameter capable to
reproduce the bending of the Fermi surface is negative in sharp contrast
with the one predicted by band calculations\cite{Duffy:1995}. Indeed, in the
context of a simple $t$-$t^{\prime }$-$U$ model a negative sign for the bare 
$t^{\prime }$ parameter is the only way to obtain a hole-like bending.

It is interesting to notice that the experimentally observed Fermi surfaces
for $YBCO$ and $NCCO$ have the same bending. This situation seems to
eliminate the possibility to describe both the electron- and hole- doped
cuprates just changing the sign of the $t^{\prime }$ parameter, at least in
the context of a simple $t$-$t^{\prime }$-$U$ model.

\begin{figure}[tb]
\centerline{\psfig{figure=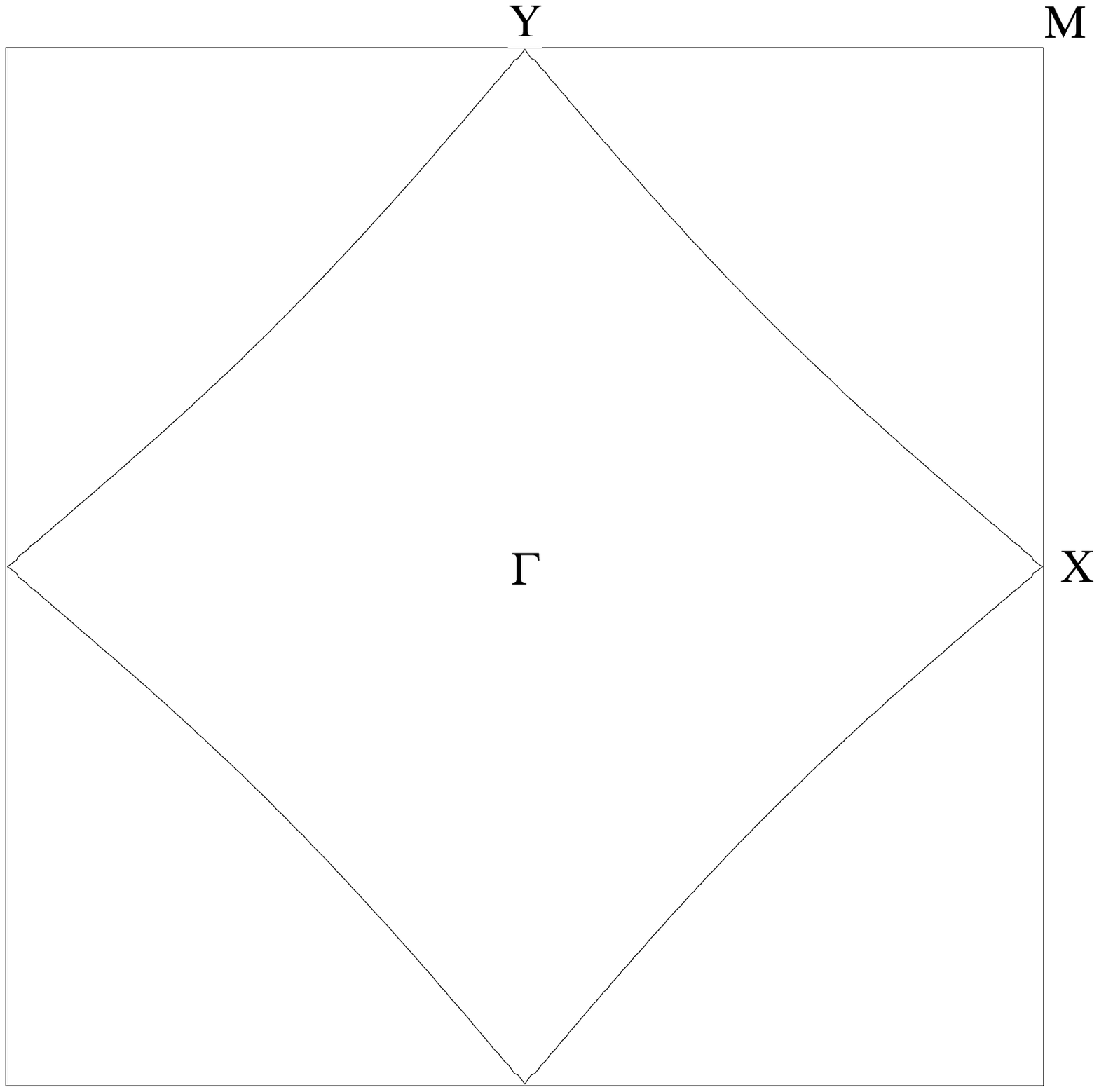,width=8cm,bbllx=16pt,bblly=143pt,bburx=579pt,bbury=706pt,clip=}}
\caption{FS for $t^{\prime }=-0.1$, $T=0.01$, $n=0.7$ and $U=3$.}
\label{LSCO2}
\end{figure}

In the case of $LSCO$, the same set of parameters already used to
successfully describe the band dispersions allows us to reproduce both the
pseudo-nesting and the hole-like bending of the Fermi surface as found by
the positron annihilation experiments and {\em LDA} calculations\cite
{Howell:1994,Xu:1987}, see Fig.~\ref{LSCO2}.

\section{Conclusions}

\label{conclusions}

Since the discovery of high-$T_c$ superconductivity, there has been a great
deal of discussion about the choice of an effective model suitable to
describe the properties of the copper-oxide superconductors. Extensive
studies of the magnetic properties, showing one spin degree of freedom in
the $Cu$-$O$ plane, have resulted in considerable evidence that
high-temperature superconductors may be modelled by an effective single-band
model. In this line of thinking, one of the most studied model is the
single-band Hubbard model. The addition of a finite $t^{\prime }$ diagonal
hopping term, that appears to be material dependent for high-$T_c$ cuprate
superconductors, has often been suggested to handle the complexity of the
experimental situation for the cuprates.

According to this, we have studied the two-dimensional $t$-$t^{\prime }$-$U$
model, by means of the Composite Operator Method. Using relations containing
the Pauli principle, we have been able to fix the dynamics in a fully
self-consistent way. Furthermore, the recovery of the Pauli principle has
assured us to satisfy the hole-particle symmetry too.

Nowadays, the experimental situation for many physical properties of cuprate
high-$T_c$ superconductors is well established. This imposes strong
constraints on the theoretical models and/or adopted approximation schemes.
The band dispersions and the Fermi surface of a large series of materials
are today well-known. {\em ARPES} data give as main information the presence
of a well-defined Fermi surface. This result gives some support to the Fermi
liquid scenario and puts some doubts on the necessity of introducing more
exotic theories.

We have computed the structure of the energy bands, the shape of the Fermi
surface and the relative position of the van Hove singularity. The
comparison with experimental data has shown that theis capable to describe
both $La_{2-x}Sr_xCuO_4$ and $Nd_{2-x}Ce_xCuO_4$, that share the property to
be $1$-layer cuprates. On the contrary, it does not seem the case for $%
YBa_2Cu_3O_{7-\delta }$ that is a $2$-layer cuprate. This can be read as a
clear signal that two-dimensional Hubbard-like models can play an important
role in describing the physics of the $1$-layer cuprates superconductors,
but that the multi-layer ones need some more complex models.

The value of the $t^{\prime }$ parameter that has been found to be necessary
for describing the electron-doped Neodymium compound is negative in sharp
contrast with the band calculations. This questions the possibility to
derive the value of the bare parameters of the model from band calculations,
as it has been done for the value of the $t^{\prime }$ parameter since its
proposal.

In conclusion, the $t$-$t^{\prime }$-$U$ model emerges as a minimal model
for $1$-layer cuprate materials.

\end{document}